\documentclass[prd,amsmath,amssymb,superscriptaddress,floatfix,nofootinbib,10pt]{revtex4}
\usepackage{orcidlink}
\usepackage{newtxtext,newtxmath}
\usepackage{amssymb,amsbsy,amsmath,amsfonts}
\usepackage{graphicx}
\usepackage{float}
\usepackage{color}
\usepackage{morefloats}
\usepackage{rotating}
\usepackage{srcltx}
\usepackage{slashed}
\usepackage{multirow}
\usepackage{verbatim}
\usepackage{hyperref}
\hypersetup{colorlinks}
\usepackage{tabularx}
\usepackage{bm}
\allowdisplaybreaks[4]

\usepackage{bbding}
\usepackage{threeparttable}
\usepackage{soul}

\DeclareUnicodeCharacter{2212}{\ensuremath{-}}

\newcommand{\PreserveBackslash}[1]{\let\temp=\\#1\let\\=\temp}
\newcolumntype{C}[1]{>{\PreserveBackslash\centering}p{#1}}
\newcolumntype{R}[1]{>{\PreserveBackslash\raggedleft}p{#1}}
\newcolumntype{L}[1]{>{\PreserveBackslash\raggedright}p{#1}}

\newcommand{\minv}{M_\text{inv}}

\newcommand{\be}{\begin{equation}}
\newcommand{\ee}{\end{equation}}

\begin{document}

\title{ \boldmath On the determination of the $D$ meson width in the nuclear medium with the transparency ratio}

\author{Victor Montesinos\orcidlink{0000-0002-6186-2777}}
\email[]{vicmonte@ific.uv.es}
\affiliation{Departamento de Física Teórica and IFIC, Centro Mixto Universidad de Valencia-CSIC Institutos de Investigación de Paterna, 46071 Valencia, Spain}

\author{Natsumi Ikeno\orcidlink{0000-0002-2334-8179}}
\email[]{ikeno@tottori-u.ac.jp}
\affiliation{Department of Agricultural, Life and Environmental Sciences, Tottori University,
Tottori 680-8551, Japan}

\author{ Eulogio Oset\orcidlink{0000-0002-4462-7919}}
\email[]{oset@ific.uv.es}
\affiliation{Departamento de Física Teórica and IFIC, Centro Mixto Universidad de Valencia-CSIC Institutos de Investigación de Paterna, 46071 Valencia, Spain}

\author{Miguel Albaladejo\orcidlink{0000-0001-7340-9235}}
\email[]{ Miguel.Albaladejo@ific.uv.es}
\affiliation{Departamento de Física Teórica and IFIC, Centro Mixto Universidad de Valencia-CSIC Institutos de Investigación de Paterna, 46071 Valencia, Spain}

\author{Juan Nieves\orcidlink{0000-0002-2518-4606}}
\email[]{ jmnieves@ific.uv.es}
\affiliation{Departamento de Física Teórica and IFIC, Centro Mixto Universidad de Valencia-CSIC Institutos de Investigación de Paterna, 46071 Valencia, Spain}

\author{Laura Tolos\orcidlink{0000-0003-2304-7496}}
\email[]{tolos@ice.csic.es}
\affiliation{Institute of Space Sciences (ICE, CSIC), Campus UAB,  Carrer de Can Magrans, 08193 Barcelona, Spain}
\affiliation{Institut d'Estudis Espacials de Catalunya (IEEC), 08860 Castelldefels (Barcelona), Spain}
\affiliation{Frankfurt Institute for Advanced Studies, Ruth-Moufang-Str. 1, 60438 Frankfurt am Main, Germany}

\begin{abstract}
We have studied the feasibility of the experimental determination of the width of a $D$ meson in a nuclear medium by using the method of the nuclear transparency. The cross section for inclusive production of a $D^+$ in different nuclei is evaluated, taking care of the $D^+$ absorption in the nucleus, or equivalently, the survival probability of the $D^+$ in its way out of the nucleus from the point of production. We use present values of the in medium width of $D$ mesons and calculate ratios of the cross sections for different nuclei to the $^{12} \text{C} $ nucleus as reference. We find ratios of the order of $0.6$ for heavy nuclei, a large deviation from unity, which indicates that the method proposed is adequate to measure this relevant magnitude, so far only known theoretically. 
\end{abstract}

\maketitle

\section{Introduction}
The study of the modification of meson properties inside a nuclear medium has attracted and continues attracting much attention in the physics community. Reviews on the subject can be found in Refs.\,\cite{Oller:2000ma,Post:2003hu,Hayano:2008vn,Metag:2017yuh,Tolos:2020aln} (see also extended information in Ref.\,\cite{Ericson:1988gk}). For light mesons there have been many methods to study the meson nucleus interaction, like scattering of pions and kaons with nuclei, or mesonic atoms, because of the availability of pion and kaon beams \cite{Oller:2000ma,Friedman:2007zza}. The experimental investigation of meson properties for which there are no beams is certainly more difficult, as is the case of $\eta$ or $\eta'$ nucleus interaction and the possibility of having  $\eta$ or $\eta'$ nuclear bound states, which have nevertheless had much experimental attention  \cite{CBELSATAPS:2013waf,CBELSATAPS:2018sck,n-PRiMESuper-FRS:2016vbn,LEPS2BGOegg:2020cth,Bass:2021rch,Xie:2016zhs,Xie:2018aeg,Ikeno:2017xyb,Skurzok:2018paa,Ikeno:2024slo}.

Our purpose here is to present an experimental method to study the modification of the $D$ meson properties in a nuclear medium. The medium properties of $D$ mesons have been thoroughly studied theoretically \cite{Tolos:2004yg,Tolos:2005ft,Mizutani:2006vq,Tolos:2007vh,Molina:2009zeg,Tolos:2009nn,Tolos:2013kva,Tolos:2013gta,Sasaki:2014asa,ReddyP:2017tgo,Buchheim:2018kss,Albaladejo:2021cxj,Montesinos:2023qbx,Montesinos:2024uhq}. However, how to observe experimentally the theoretical predictions remains still an unexplored field. In the present work we propose a method which is based on the use of the transparency ratio in $\gamma$-nucleus collisions producing $D$ mesons. 
  
The method of the transparency ratio was proposed in Ref.\,\cite{Hernandez:1992rv} to determine the width of antiprotons in a nuclear medium. While there are $\bar p$ beams available, the measurement of the $\bar p$ absorption in nuclei was found basically useless, because with such a large $\bar p p$ annihilation, the $\bar p$ absorption cross section is roughly $\pi R^2$, with $R$ the radius of the nucleus, independently of the $\bar p$ width in the nucleus, meaning that all $\bar p$ will be basically absorbed. If one reduces this width, the $\bar p$ are absorbed later in the nucleus but are absorbed anyway. The insensitivity of these experiments to determine the $\bar p$ properties in the medium stimulated the suggestion of using the production of $\bar p$ in nuclei rather than the absorption from a $\bar p$ beam. The ideal reaction is to use photons that can reach all parts of the nucleus with negligible absorption and then produce the antiprotons and see their survival probability. This magnitude is then very sensitive to the $\bar p$ width in the medium and then, by measuring the $\bar p$ production in different nuclei, one can make a determination of the $\bar p$ width in the medium. 
 
The method has become popular and has been used with the same purpose to determine particle properties in the medium \cite{Magas:2004eb, Kaskulov:2006zc, Garrow:2001di, Lava:2004zi, CBELSATAPS:2012few, CLAS:2012usg,Tolos:2010fq,Ishikawa:2004id,Cabrera:2003wb}. in Ref.\,\cite{CBELSATAPS:2012few} the transparency ratio method was used in the $\gamma A \to \eta' A'$ reaction comparing the $\eta'$ production in several nuclei,  $^{12}$C, $^{40}$Ca, $^{93}$Nb and $^{208}$Pb, and to eliminate the possible contribution of multinucleon production processes the production rates were normalized to that of the $^{12}$C. We shall investigate the same nuclei here and study the $\gamma A \to D^+ D^- A'$ reaction in these different nuclei to see the sensitivity to the $D^+$ width in the medium.  We shall take the results for the $D$ medium  properties from \cite{Tolos:2009nn}. The work will be exploratory and we shall make some approximations, which, however, still allow to obtain reasonable transparency ratios. When the idea proposed here is eventually implemented experimentally, some improvements can be done which we will mention at the end.

This paper is organized as follows. In Sec.~\ref{sec:formalism} we introduce the formalism of the transparency ratio method applied to the $\gamma A \to D^+ D^- A^\prime$ reaction, starting by presenting a simple model for the absorption of the $D^+$ meson produced inside the nucleus, and then paying attention to the kinematics of the three-body final state that enter the phase space integral needed for the computation of the nucleus cross section. Sec.~\ref{sec:Results} presents the theoretical predictions for the transparency ratio of the $D^+$ meson. %
Since the $D^+$ width is not exactly determined in the model of Ref.\,\cite{Tolos:2009nn} at the $D^+$ momenta involved in this work, we consider different scenarios for the $D^+$ width inspired by {the} said model. 
We also engage in a discussion about the optimal $\gamma$ energy needed for the production of a maximal number of $D^+$ mesons with small momentum. We finish this section considering some caveats that the computation and measurement present, giving as well some possible solutions that could be taken into consideration in future works. {In Sec.~\ref{sec:improved} we present an improved method where realistic nuclear densities are used and aspects of the $t$ dependence of the $D$ production amplitudes are considered.} Finally, in Sec.~\ref{sec:Conclusions} we present our conclusions.

\section{Formalism}\label{sec:formalism}
\begin{figure}[t]
    \centering
    \includegraphics{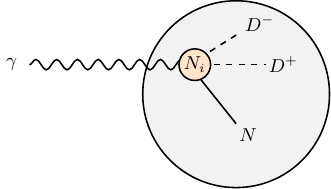}
    \caption{Schematic picture of the process considered for the $\gamma A \to D^+ D^- A'$ reaction.}
    \label{fig:reaction}
\end{figure}

With the prospective character of the paper we start from using a simplified picture for the nuclei, assuming a sphere of constant density $\rho_0=0.16$ fm$^{-3}$.  {In Sec.~\ref{sec:improved}} we shall use realistic densities. Then,
\be
\rho(r) = 
\left\{
\begin{aligned}
    &\rho_0 , & |\vec{ r} | &\le R \\
    &0 , &|\vec{ r} | &> R 
\end{aligned}
\right.
; \quad R= \left(\frac{3 A}{4\pi \rho_0}\right)^{\frac 1 3},
\ee
where $R$ is the nuclear radius and $A$ is the nuclear mass number. Our picture for $D^+$ production will start from an elementary reaction
\be\label{e:reaction}
\gamma N_i \to N D^+ D^-,
\ee
where $N_i$ and $N$ refer to the initial and final nucleons, respectively. Note that we must produce a $D^-$ in connection with $D^+$ to conserve charm. If we were looking at $D^-$ production we could have it through a two-body final state as $\gamma N \to D^- (\Lambda_c,\, \Sigma_c,\, \dots)$, but in the absence of elementary baryons with $\bar c$, we stick to the $D^+$ production through the mechanism of Eq.~\eqref{e:reaction}. The process proceeds as depicted in Fig.~\ref{fig:reaction}.
We are not concerned with what happens to the $D^-$ and the nucleon, and only follow the $D^+$ and evaluate the survival probability to arrive at the surface of the nucleus and leave it to be detected.

Following Ref.~\cite{Tolos:2009nn} we use the selfenergy of the $D^+$ in the nuclear medium, as a function of the nuclear density and its momentum in the rest frame of the nucleus. We know that
\be\label{e:absortionunitl}
dP=-\frac{\text{Im} \ \Pi}{p_{D,\text{lab}}}dl
\ee
is the differential probability of $D^+$ absorption in a distance $dl$. In Eq.~\eqref{e:absortionunitl} $\Pi$ stands for the $D^+$ selfenergy in the nuclear medium, which includes inelastic and quasielastic channels; and $p_{D,\text{lab}}$ is the modulus of the three-momentum of the $D^+$ in the laboratory (lab) frame, where the nucleus is at rest. Then, the survival probability $S(\vec{ r}, \vec{ p}_{D,\text{lab}})$ of this $D^+$ without undergoing reactions is given by
\be\label{e:survprob}
S(\vec{ r}, \vec{ p}_{D,\text{lab}}) = \text{exp}\left[-\int_0^\infty dl \ \frac{- \text{Im} \ \Pi\left(\rho(\vec{ r}{}^{\,\prime}),\vec{ p}_{D,\text{lab}}\right)}{p_{D,\text{lab}}}\right],
\ee
with
\be
\vec{ r}{}^{\,\prime} = \vec{ r} + l \ \frac{\vec{ p}_{D,\text{lab}}}{p_{D,\text{lab}}}.
\ee
For the constant density nuclei, this integral can be calculated analytically and it is given by
\be
S(\vec{ r}, \vec{ p}_{D,\text{lab}})  = \text{exp}\left[\frac{\text{Im} \ \Pi\left(\rho_0,\vec{ p}_{D,\text{lab}}\right)}{p_{D,\text{lab}}} L(r,\theta^\prime)\right] ,
\ee
with
\be\label{e:analyticL}
L(r,\theta^\prime) = \sqrt{R^2-r^2\sin^2 \theta^\prime}-r\cos \theta^\prime,
\ee
where $\theta^\prime$ is the angle between $\vec{r}$ and $\vec{p}_{D,\text{lab}}$ (in the nucleus rest frame).

The next step is to evaluate the phase space for the process and the cross section for each nucleus. We take the photon in the $z$ direction and first evaluate the phase space in the $\gamma N_i$ center of mass frame, assuming the $N_i$ momentum ($N_i$ being the initial nucleon) to be at rest in the nucleus rest frame, which is a good assumption given the large momenta of the photons involved in the reaction. Then, the cross section is given by
\be\label{e:CrossSection}
\sigma = \frac{m_N^2}{(s-m_N^2) \sqrt s} \frac{1}{32\pi^4} \int_{\minv^\text{min}}^{\minv^\text{max}} p p^\prime d\minv \int_{-1}^1 d\cos\theta \int_0^{2\pi} d\phi \ |t|^2 2\pi \int_0^R dr \ \rho_0 r^2 \int_{-1}^1 d\cos\tilde\theta \ S(\vec{ r}, \vec{ p}_{D,\text{lab}}).
\ee
where $|t|^2$ is the squared modulus of the $\gamma N_i \to D^+D^- N$ transition matrix, which we take as a constant in the simplifying path that we follow and then it will cancel out when evaluating the transparency ratio at the end. {This will be improved in Sec.~\ref{sec:improved}.} In Eq.~\eqref{e:CrossSection}, $\vec p_\gamma$ is given by
\begin{align}\label{e:pgamma}
    \vec p_\gamma &= p_\gamma
    \begin{pmatrix}
        0\\
        0\\
        1
    \end{pmatrix}
    , &
    p_\gamma &= \frac{1}{2 \sqrt s}(s-m_N^2) \quad \text{(in} \ \gamma N_i \ \text{center \ of \ mass \ frame)}, &
    p_{\gamma,\text{lab}} &= \frac{1}{2 m_N}(s-m_N^2).
\end{align}
Additionally, $p$ is the momentum of the $D^+$ in the $\gamma N_i$ center of mass frame
\begin{align}
    \vec p &= p
    \begin{pmatrix}
        \sin\theta \cos\phi\\
        \sin\theta \sin\phi\\
        \cos\theta
    \end{pmatrix}
    , &
    p &= \sqrt{\omega^2(p)-m_D^2} , &
    \omega(p) &= \frac{s + m_D^2 - \minv^2}{2 \sqrt s},
\end{align}
and $\minv$ is an integration variable corresponding to the invariant mass of $D^-N$ (with $N$ the emitted nucleon in $\gamma N_i \to D^+ D^- N$).
The integration limits of $\minv$ are
\be
\begin{aligned}
    \minv^\text{min}&= m_{D^-} + m_N, \\
    \minv^\text{max}&= \sqrt s - m_{D^+}.
\end{aligned}
\ee
Furthermore, $p^\prime$ denotes the momentum of the $D^-$ in the $D^-N$ rest frame
\be\label{e:pprime}
p^\prime = \frac{\lambda^{\frac12}\left(\minv^2, m_{D^-}^2,m_N^2\right)}{2 \minv},
\ee
with $\lambda$ the K\"all\'en function, and $\vec r$ is the position inside the nucleus where the $\gamma N_i \to D^+ D^- N$ reaction takes place, given by
\be
\vec r = r 
\begin{pmatrix}
    \sin\tilde\theta \cos\tilde\phi \\
    \sin\tilde\theta \sin\tilde\phi \\
    \cos\tilde\theta
\end{pmatrix}
= r 
\begin{pmatrix}
    \sin\tilde\theta  \\
    0 \\
    \cos\tilde\theta
\end{pmatrix}
.
\ee
In the previous expression for $\vec r$ we take $\tilde \phi=0$ because when doing the integral over all angles of $\vec p$ it is unnecessary to make an extra $\tilde \phi$ rotation in $\vec r$.

This settles the issue of the phase space in the evaluation of $\sigma$ and the only step missing is how to calculate the momentum of the $p_{D^+}$ in the lab frame of the nucleus at rest. This is done by means of a boost from the frame where the initial nucleon $N_i$ has a momentum
\be
\vec p_{N_i} = - \vec p_\gamma
\ee
to the frame where the $N_i$ is at rest. We apply the general formula given in Ref.~\cite{FernandezdeCordoba:1993az}
\be
\vec p_{D,\text{lab}} = \left[\left(\frac{E(p_i)}{m_N}-1\right) \frac{\vec p \cdot \vec p_i}{|\vec p_i|^2} - \frac{\omega(p)}{m_N}\right] \vec p_i + \vec p ,
\ee
where $\vec p_i \equiv \vec p_{N_i}$ and $E(p_i)$ are the three-momentum and energy of the initial nucleon in the $N_i \gamma$ center of mass frame.

Once this is done, one defines the transparency ratio as
\be
\tilde T = \frac{\sigma_A}{A \ \sigma_N} ,
\ee
where $\sigma_N$ is the nucleon cross section, $A$ is the nucleus mass number and $\sigma_A$ is the corresponding nucleus cross section. However, we prefer to normalize to the $^{12}$C results in order to reduce the effect of multi-step processes leading to the same final states, as was done in Ref.~\cite{CBELSATAPS:2012few}, and finally define
\be\label{e:TR}
T = \frac{\sigma_A/A}{\sigma_{12}/12} = \frac{12}{A} \frac{\sigma_A}{\sigma_{12}}.
\ee

Next we do some considerations on the $\gamma$ energies needed for the process. The minimum value of $\sqrt s$ for $\gamma N_i$ is
\be
(\sqrt s){}_\text{min} = m_N + m_{D^-} + m_{D^+} \simeq 4.678 \ \text{GeV},
\ee
which corresponds to a $p_{\gamma,lab} \simeq 11.17$ GeV. We want to optimize the original set up to the input that we have. The medium selfenergy of the $D$ meson is calculated in Ref.~\cite{Tolos:2009nn} for momenta up to about $1$ GeV. Then it is easy to calculate the kinematics of the emitted $D^+$ and see that if we have small values of $\sqrt s$, the minimum momenta of the $D^+$ in the lab frame are of the order of $3$ GeV. But we also see that if we take values of around $\sqrt s \simeq 10$ GeV or higher, then the minimum $p_{D,\text{lab}}$ values stabilize around $1.4\sim1.5$ GeV.

There is another thing we do to maximize the efficiency of the proposed experiment. Because our calculations are done up to about $p_{D,\text{lab}}=1$ GeV, and the minimum momenta are also bigger than $1.4$ GeV, we select momenta $p_{D,\text{lab}}<2$ GeV. Then, we check which is the photon energy which maximizes the fraction of $D^+$ that come with a momentum in that range.

\section{Results}\label{sec:Results}
Following Ref.~\cite{Tolos:2009nn}, we take
\be\label{e:ImPi}
- \text{Im}\ \Pi\left(\rho_0,p_{D,\text{lab}}=1 \ \text{GeV}\right) = 0.1 \ \text{GeV}^2,
\ee
which, by using $\Gamma_D = - \text{Im} \ \Pi / E_D$, gives a $D$ width in the medium of about $47$ MeV for $p_{D,\text{lab}}=1$ GeV. There is no need to worry about the density dependence at this point since our nuclei have constant density $\rho_0$.

\begin{figure}[t]
    \centering
    \includegraphics[width=0.5\textwidth]{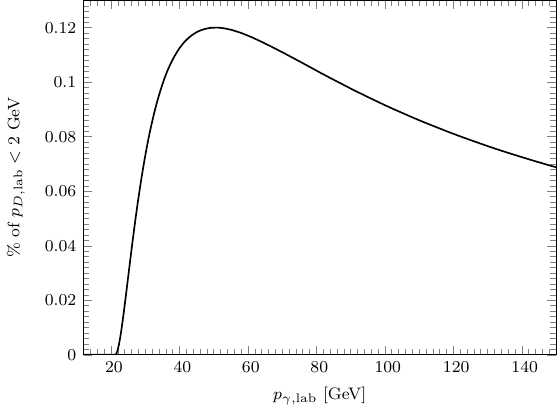}
    \caption{Percentage of $D^+$ mesons with $p_{D,\text{lab}}<2$ GeV as a function of $\sqrt s$ of $\gamma N_i$.}
    \label{fig:FractionMomentum}
\end{figure}

In Fig.~\ref{fig:FractionMomentum} we show the fraction of $D^+$ momenta with $p_{D,\text{lab}}<2$ GeV as a function of the photon three-momentum in the lab frame. This is computed using a Monte Carlo integration method for the phase space of Eq.~\eqref{e:CrossSection} and counting the number of times that the $p_{D,\text{lab}}$ is smaller than $2$ GeV with respect to the total number of Monte Carlo points in the integral. As we can see from this figure, the optimal photon momentum is of about $p_{\gamma,\text{lab}}=50$ GeV. This $p_{\gamma,\text{lab}}$ corresponds to a photon nucleon invariant mass of about $\sqrt s = 10$ GeV. We choose this energy to carry out the calculations of the transparency ratio.

\begin{figure}[t]
    \centering
    \includegraphics[width=.5\textwidth]{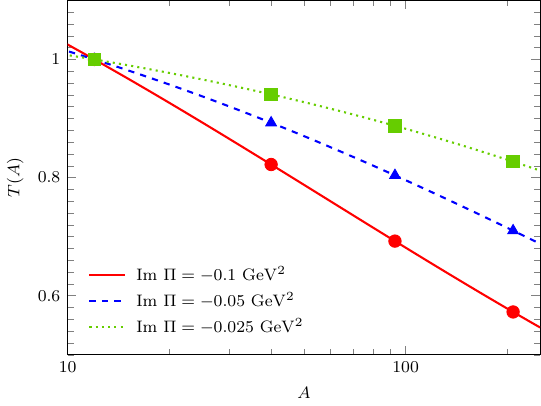}
    \caption{transparency ratio of Eq.~\eqref{e:TR}, normalized to $^{12}$C as a function of the mass number of nuclei. The three lines show the transparency ratio calculated with the different values of Im $\Pi$. The points in the figure indicate the results for $^{12}$C, $^{40}$Ca, $^{93}$Nb, and $^{208}$Pb.}
    \label{fig:TransparencyRatio}
\end{figure}

In Fig.~\ref{fig:TransparencyRatio} we show the results for the transparency ratio of Eq.~\eqref{e:TR}.
As we can see, the function $T(A)$ diverges substantially from unity, which tells us that an actual measurement of this ratio can give information on $\text{Im}\ \Pi$. This was actually the case when studying the emission of $\eta^\prime$ in nuclei \cite{CBELSATAPS:2012few}, from where one could induce, using actual data, that the width of the $\eta^\prime$ in a nuclear medium at $\rho=\rho_0$ was $\Gamma_{\eta^\prime}=-\text{Im} \ \Pi_{\eta^\prime}/E_{\eta^\prime}\simeq 15\sim 25$ MeV. To show the sensitivity of $T(A)$ to the value of $\text{Im} \ \Pi$, we redo the calculations using $\text{Im}\ \Pi = -0.05$ GeV$^2$ and $-0.025$ GeV$^2$, which correspond to a half and a quarter of the value presented in Eq.\,\eqref{e:ImPi}, respectively. We see in Fig.~\ref{fig:TransparencyRatio} that the values of $T(A)$ are now {closer to $1$}, but still significantly different from unity, such that one can still distinguish possible values of $\text{Im} \ \Pi$ smaller than what was evaluated in Ref.~\cite{Tolos:2009nn}. 

Once this point is reached, let us discuss some related caveats. Our calculations are based on the assumption that $\text{Im}\ \Pi$ is mostly due to genuine absorption, where the $D^+$ disappears and that the quasielastic contribution $D^+N \to D^+ N^\prime$ is relatively small. While this is the case in our calculations \cite{Tolos:2009nn}, where the main channels for $\text{Im}\ \Pi$ stem from $DN \to \pi \Sigma_c$, $\pi \Lambda_c$, $K \Xi_c$, $D^* N$, there is another filter that one could do in future studies when data are available. Indeed, the kinematics of the process, with very fast photons, lead to $D^+D^-N$ all in a very narrow cone in the forward direction in the $\gamma A$ rest frame. The $D^+$ that we have selected have a small momenta, as we say between $1.4\sim2$ GeV. These $D^+$ colliding with other nucleons in the nuclei with $D^+N\to D^+ N^\prime$ will show a dispersion of angles, where the final $D^+$ will not be so forward peaked as the primary $D^+$ produced. One can then put an extra filter on forward angles for the $D^+$, such that only these which do not undergo any reaction are observed. In this case we would obtain the full $\Pi$ for $D^+$ in the medium. One can argue similarly for primary $D^+$ produced with momentum larger than $2$ GeV, which lose energy after a quasielastic collision. 
In a further step one can consider the shadowing of photons at the energy where the experiment is done, which is always much smaller than the strong absorption of the $D$ mesons \cite{CBELSATAPS:2012few}.

\subsection{Improvements on nuclei density profiles and momentum-dependent photoproduction amplitudes}\label{sec:improved}
In the calculations done we have made approximations, which allow us to obtain sensible results with a minimum computing time. Yet, one might rightly state that the consideration of  actual nuclear densities and the momentum dependence of the $\gamma N_i \to D^+ D^- N$ transition amplitudes would alter the results obtained notably.  In this section we address these two issues.

In the first place we take the nuclear density $\rho(r)$ from Ref.~\cite{DEVRIES1987495}, as used in Ref.~\cite{CBELSATAPS:2012few}. For $^{12}\text{C}$ we use a harmonic oscillator shape
\begin{equation}
    \rho(r) = \tilde \rho_0 \left(1 + c \left(\frac{r}{R}\right)^2\right) e^{-\left(r/R\right)^2},
\end{equation}
with $\tilde \rho_0$ normalized such that $\int d^3 r \ \rho(r) = 12$ and $c = 1.082$, $R= 1.692$ fm. For $^{40}\text{Ca}$, $^{93}\text{Nb}$, $^{208}\text{Pb}$ we take the Woods-Saxon shape
\begin{equation}
    \rho(r) = \frac{\tilde \rho_0}{1 + \mathrm{exp}\left(\frac{r-R}{a}\right)},
\end{equation}
with $\tilde \rho_0$ normalized such that $\int d^3 r \ \rho(r) = A$ for each nucleus and the parameters $R$ and $a$ given in Table \ref{tab:WSpar}.
\begin{table}[ht]
    \centering
    \begin{tabular}{ccc}
                        Nucleus&  $R$ [fm]    & $a$ [fm]  \\ \hline
         $^{40}\text{Ca}$  & $3.51$    & $0.563$ \\
         $^{93}\text{Nb}$  & $4.87$    & $0.573$ \\
         $^{208}\text{Pb}$  & $6.62$    & $0.549$ \\
    \end{tabular}
    \caption{Values of $a$ and $R$ for $^{40}\text{Ca}$, $^{93}\text{Nb}$, $^{208}\text{Pb}$.}
    \label{tab:WSpar}
\end{table}
Now the integral $S(\vec r,\, \vec p_{D,\mathrm{lab}})$ of Eq.~\eqref{e:survprob}, which before could be done analytically by means of Eq.~\eqref{e:analyticL}, has to be done numerically for each value of $\vec r$ and $\vec p_{D,\mathrm{lab}}$ appearing in the Monte Carlo calculation. This integral is done using a Gaussian integration algorithm. $\mathrm{Im} \ \Pi$ becomes now $\mathrm{Im} \ \Pi(\tilde \rho_0) \rho(r)/\tilde \rho_0$.

The next issue is the momentum dependence of the $D^+ D^-$ photoproduction amplitude. One reason to normalize the production of different nuclei to the one of $^{12}\text{C}$ was that details on the production amplitude largely canceled in that ratio, which was essentially tied to the width of the produced particle in the medium. In the present case we can benefit to account for the momentum dependence of the amplitude from the work of Ref.~\cite{Siddikov:2023qbd}. In that paper a calculation is done for the $\gamma p \to D^+ D^{\ast -} p$ and related reactions, and momentum and angular dependencies are evaluated. While taking into account explicitly all the dependencies is beyond reach, we have taken advantage of these results to prove the large insensitivity of the transparency ratio to such details. For this purpose we have chosen the dependence of the cross section in the variable $t^\prime = (p_{N_i}-p_N)^2$, which gives rise by far to the largest changes on the cross section, a two orders of magnitude reduction from $t^\prime \simeq 0$ to $- t^\prime \simeq 1 \ \text{GeV}^2$. 

Hence, we have taken for our transition matrix element squared $|t|^2$ the dependence of
\begin{equation}
    |t|^2 = |t_0|^2 \mathrm{exp}\left(4.6 \frac{ t^\prime}{ 1 \ \text{GeV}^2}\right).
\end{equation}
Taking this into consideration forces us to introduce more variables into the evaluation of the cross section of Eq.~\eqref{e:CrossSection}. Indeed, in this formula the $\vec p^\prime$ integration of the momentum of the $D^-$ in the $D^- N$ rest frame was done as
\begin{equation}
    \int d^3p^\prime = \int p^{\prime 2} d p^\prime d\tilde \Omega(p^\prime) = 4\pi \int p^{\prime 2} dp^\prime.
\end{equation}
The value of $p^\prime$ is still given by Eq.~\eqref{e:pprime} but now we replace
\begin{equation}
    4\pi \to \int_{-1}^{+1}d\mathrm{cos}\tilde \theta^\prime \int_0^{2\pi} d\tilde \phi^\prime
\end{equation}
such that
\begin{equation}
    \vec p^\prime = p^\prime
    \begin{pmatrix}
        \sin\tilde \theta^\prime \cos\tilde \phi^\prime\\
        \sin\tilde \theta^\prime \sin\tilde \phi^\prime\\
        \cos\tilde \theta^\prime
    \end{pmatrix}.
\end{equation}

To evaluate the nucleon momentum transfer we boost the initial nucleon momentum from the $\gamma N_i$ rest frame, where $\vec p_{N_i} = -\vec p_{\gamma}$ with $\vec p_\gamma$ given in Eq.~\eqref{e:pgamma}, and where the $D^- N$ system has momentum $-\vec p$, to the frame where the $D^- N$ system is at rest. We find now
\begin{equation}
    \vec p_{N_i,\text{boost}} =-\left[\left(\frac{E_{D^-N}(\vec p^2)}{M_\text{inv}}-1\right) \frac{\vec p \cdot \vec p_\gamma}{\vec p^2} - \frac{E_{N_i}(\vec p_\gamma^2)}{M_\text{inv}}\right] \vec p - \vec p_\gamma,
\end{equation}
where $E_{D^-N}(\vec p^2) = \sqrt{M_\text{inv}^2 + \vec p^2}$  and $E_{N_i}(\vec p_\gamma^2) = \sqrt{m_N^2 + \vec p_\gamma^2}$ are the energy of the $D^- N$ system  and the initial nucleon in the $\gamma N_i$ rest frame, respectively. Now we can evaluate the momentum transfer squared:
\begin{equation}
    \left(p_{N_i}^2 - p_N^2\right) = 2 m_N^2 - 2 E_{N_i,\text{boost}} \sqrt{m_N^2 + \vec p^{\prime 2}} - 2 p_{N_i,\text{boost}}\cdot \vec p^\prime,
\end{equation}
where $E_{N_i,\text{boost}}=\sqrt{m_N^2 + \vec p_{N_i,\text{boost}}^2}$. We have introduced two new integrals in the Monte Carlo integration, plus the external Gaussian integration for the computation of the $S(\vec r,\, \vec p_{D,\text{lab}})$ quantity. This yields an increase in the computing time, but the calculation remains yet perfectly feasible.

\begin{figure}
    \centering
    \includegraphics[width=0.5\linewidth]{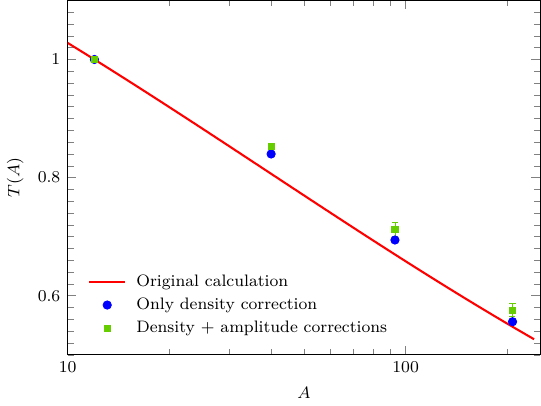}
    \caption{Green squares: results for the transparency ratio to $^{12}\text{C}$ using realistic densities with $\mathrm{Im} \ \Pi = \mathrm{Im}\ \Pi(\tilde \rho_0) \rho(r)/\tilde \rho_0$ and $\mathrm{Im} \ \Pi(\tilde \rho_0)= -0.1$ GeV$^2$, as well as the momentum dependence of the $D^+D^-$ photoproduction amplitude. Red solid line: result of Fig.~\ref{fig:TransparencyRatio} for $\mathrm{Im}\ \Pi =-0.1$ GeV$^2$ with the simplified treatment. Blue circles: results obtained when only the changes in the nuclear density are considered.}
    \label{fig:newTR}
\end{figure}

In Fig.~\ref{fig:newTR} we show the new results that we obtain for the transparency ratio, after normalizing to the $^{12}\text{C}$ results, compared to those obtained before with the simplified formulas. There are some changes, but the results are very similar. For instance, for $^{208}\text{Pb}$ the differences are of the order of $5 \%$, and a bit bigger for $^{40}\text{Ca}$ and $^{93}\text{Nb}$. We should expect fewer changes from other momentum dependencies of the $\gamma N_i \to D^+ D^- N$ amplitude. We confirm what was found in \cite{CBELSATAPS:2012few} about the insensitivity of the transparency ratio to details of the production. We have checked that most of the differences in Fig.~\ref{fig:newTR} are indeed largely due to the use of the realistic densities, while the details of the transition amplitude are relatively unimportant. The errors for the final results {(green squares in Fig. \ref{fig:newTR})} are obtained by running several times the Monte Carlo calculations with different seeds.  We take the average from the different runs and the dispersion of the results with respect to the average gives us the error. {In the original calculations (red line), as well as in the results considering only the nuclear density distributions correction (blue circles), the errors arising from the Monte Carlo integration are negligible.}

As to the actual facilities where the reactions could be done, the GlueX Facility in Jefferson Lab \cite{GlueX:2019mkq} can produce photons up to an energy around $12$ GeV. This energy, although sufficient to produce the $D^+D^-$ pair (the minimum $p_{\gamma,\text{lab}}$ needed to generate the $D^+D^-$ pair at threshold is of $11.2$ GeV), is not able to generate the desirable low-momentum $D^+$ mesons with $p_{D^+,\text{lab}}<2$ GeV. As seen in Fig.~\ref{fig:FractionMomentum}, the magnitude of $p_{\gamma,\text{lab}}$ required for this is of around $21$ GeV. However, if the energy of the electrons at JLab were to be upgraded to $22$ GeV, as has been proposed in Ref.~\cite{Accardi:2023chb}, it would be possible to produce low-momentum $D^+$ mesons, although with small statistics. Apart from GlueX, there are other future accelerators that will be able to generate the energies needed for $\gamma A \to A^\prime D^+ D^-$ process to occur while maximizing the number of low-momentum $D^+$ mesons. One such example is the EIC in the US, which will reach electron-proton center of mass energies of up to $100$ GeV \cite{AbdulKhalek:2021gbh}. Another example is the EicC in China, where they will collide electrons and several nuclei for center of mass energies ranging from $10$ to $15$ GeV \cite{Anderle:2021wcy}. The method proposed to measure the width of the $D$ in a nuclear medium is to our knowledge the first experimental reaction suggested to actually measure such physical magnitude.

\section{Conclusions}\label{sec:Conclusions}
We have explored the feasibility of measuring the width of the $D$ mesons in nuclei by using the method of the nuclear transparency. The reaction suggested is $\gamma A \to D^+ D^- A'$, where the elementary reaction taking place in the nucleus is $\gamma N_i \to D^+ D^- N$, with $N_i$ the nucleons in the nucleus. The $D^+$ is followed once produced, and the probability that it survives without being absorbed in the nucleus is evaluated. Then the total cross section for the nuclear reaction is calculated, with a choice of variables in the integration that allows us to easily evaluate different magnitudes of relevance for the reaction, like the fraction of $D$ mesons that leave the nucleus with a certain momentum, the angles of the produced particles, invariant masses of the $D^-$ and $N$ particles not observed, etc. 

We perform the calculations using the results for $\text{Im}\ \Pi$ in the nuclear medium of Ref.~\cite{Tolos:2009nn}, and calculate the ratio of cross sections for $D^+$ production in  different nuclei to that of the $^{12}$C nucleus. Since $\text{Im}\ \Pi$ is calculated up to about $1$ GeV of $D$ momentum, and there is a minimum momentum of about $1.4$ GeV for the $D$ produced, we put a filter of $D^+$ observed with momentum smaller that $2$ GeV.  To optimize the efficiency of the reaction we look for the energy of the photons for which the fraction of $D$ mesons produced in this range is maximum, which we obtain around  $\sqrt s$  for $\gamma N_i$ for a nucleon $N_i$ of the nucleus of about $10$ GeV. The transparency ratios that we get relative to the $^{12}$C  nucleus reach values of $0.6$ for nuclei like $^{208}$Pb, which is a substantial reduction which should be easy to see in actual experiments. The reactions could be carried in a future experiment like the EIC or the EicC, or in a future upgrade of GlueX. 

We have made some approximations to make the calculations easier and clearer in this first exploratory approach. With respect to these approximations, we have also considered a list of issues where improvements can be done, once one knows the photon energies that will be used and other particular circumstances in an eventual implementation of the experiment. For the moment we have demonstrated the feasibility of the experiment and the optimal conditions for a successful determination of the $D^+$ selfenergy evaluated theoretically. We hope the work done stimulates interest in the experimental side. 

\acknowledgments
This work was supported under contracts No.\,PID2020-112777GB-I00, and No.\,PID2022-139427NB-I00 financed by the Spanish MCIN/AEI/10.13039/501100011033/FEDER, UE, by Generalitat Valenciana under contract PROMETEO/2020/023, and from the project CEX2020-001058-M (Unidad de Excelencia ``Mar\'{\i}a de Maeztu'') and CEX2023-001292-S (Centro de Excelencia ``Severo Ochoa''). This project has received funding from the European Union Horizon 2020 research and innovation programme under the program H2020-INFRAIA-2018-1, grant agreement No.\,824093 of the STRONG-2020 project.  This work of N. I. was partly supported by JSPS KAKENHI Grant Number 24K07020. M.\,A. and V.\,M.~are supported through Generalitat Valenciana (GVA) Grants No.\,CIDEGENT/2020/002 and ACIF/2021/290, respectively. L.\,T. also acknowledges support from the CRC-TR 211 'Strong-interaction matter under extreme conditions'- project Nr. 315477589 - TRR 211 and from the Generalitat de Catalunya under contract 2021 SGR 00171.

\bibliographystyle{JHEP}
\bibliography{references.bib}
\end{document}